# Title: The solar nebula origin of (486958) Arrokoth, a primordial contact binary in the Kuiper belt

**Authors:** W. B. McKinnon[1]*, D. C. Richardson[2], J. C. Marohnic[2], J. T. Keane[3], W. M. Grundy[4,5], D. P. Hamilton[2], D. Nesvorný[6], O. M. Umurhan[7,8], T. R. Lauer[9], K. N. Singer[6], S. A. Stern[6], H. A. Weaver[10], J. R. Spencer[11], M. W. Buie[6], J. M. Moore[7], J. J. Kavelaars[11], C. M. Lisse[10], X. Mao[1], A. H. Parker[6], S. B. Porter[6], M. R. Showalter[8], C. B. Olkin[6], D. P. Cruikshank[7], H. A. Elliott[12,13], G. R. Gladstone[12], J. Wm. Parker[6], A. J. Verbiscer[14], L. A. Young[6], and the New Horizons Science Team †.

**Affiliations:**

[1]Department of Earth and Planetary Sciences and McDonnell Center for the Space Sciences, Washington University in St. Louis, St. Louis, MO 63130, USA.

[2]Department of Astronomy, University of Maryland, College Park, MD 20742, USA.

[3]Division of Geological and Planetary Sciences, California Institute of Technology, Pasadena, CA 91125, USA.

[4]Lowell Observatory, Flagstaff, AZ 86001, USA.

[5]Northern Arizona University, Flagstaff, AZ 86011, USA

[6]Southwest Research Institute, Boulder, CO 80302, USA.

[7]NASA Ames Research Center, Space Science Division, Moffett Field, CA 94035, USA.

[8]SETI Institute, Mountain View, CA 94043, USA.

[9]NSF's National Optical-Infrared Astronomy Research Laboratory, Tucson, AZ 85726, USA.

[10]Johns Hopkins University Applied Physics Laboratory, Laurel, MD 20723, USA.

[11]National Research Council of Canada, Victoria, BC V9E 2E7, Canada

[12]Southwest Research Institute, San Antonio, TX 78238, USA.

[13]Department of Physics and Astronomy, University of Texas, San Antonio, TX 78249, USA.

[14]Department of Astronomy, University of Virginia, Charlottesville, VA 22904, USA.

*Correspondence to: mckinnon@wustl.edu

† Team members and affiliations are listed in the supplementary materials.

**Abstract:** The *New Horizons* spacecraft's encounter with the cold classical Kuiper belt object (486958) Arrokoth (formerly 2014 MU$_{69}$) revealed a contact-binary planetesimal. We investigate how it formed, finding it is the product of a gentle, low-speed merger in the early Solar System. Its two lenticular lobes suggest low-velocity accumulation of numerous smaller planetesimals within a gravitationally collapsing, solid particle cloud. The geometric alignment of the lobes indicates the lobes were a co-orbiting binary that experienced angular momentum loss and subsequent merger, possibly due to dynamical friction and collisions within the cloud or later gas drag. Arrokoth's contact-binary shape was preserved by the benign dynamical and collisional environment of the cold classical Kuiper belt, and so informs the accretion processes that operated in the early Solar System.

**Main Text:** Following its encounter with Pluto in 2015 (*1*), the *New Horizons* spacecraft continued further into the Kuiper belt (*2*). This included a flyby of (486958) Arrokoth (also informally known as Ultima Thule), discovered in a dedicated *Hubble Space Telescope* campaign (*3*). Arrokoth's orbit has a semimajor axis $a_\odot$ = 44.2 astronomical units (au), eccentricity $e = 0.037$, and inclination $i = 2.54°$, making it a member of the cold classical Kuiper belt (CCKB), a reservoir of mainly small bodies on dynamically cold orbits, i.e., those with low-to-moderate $e$ and low $i$ (typically $i < 5°$), in the outer Solar System (*4*). CCKB objects have a steeper size-frequency distribution, higher binary fraction, higher albedos, and redder optical colors than the dynamically hot and Neptune-resonant populations of the Kuiper belt, implying a distinct formation mechanism and/or evolutionary history (*4*). CCKB objects are thought to have formed in place and remained largely undisturbed by the migration of the Solar System's giant planets (*4, 5, 6*), making them unperturbed remnants of the original protoplanetary disk.

The encounter showed Arrokoth is a bi-lobed object, consisting of two discrete, quasi-ellipsoidal lobes (equivalent spherical diameters 15.9 and 12.9 km, respectively) joined at a narrow contact area or "neck" (Fig. 1) *(7,8)*. We interpret this geometric, co-joined object as a contact binary, i.e., two formerly separate objects that have gravitated towards each other until they touch. The larger lobe (hereafter LL) is more oblate than the smaller lobe (hereafter SL) *(8)*. Arrokoth rotates with a 15.92-hr period at an obliquity of 99° (the angle between its rotation axis and heliocentric orbital plane). The short axes of both lobes are aligned to within a few degrees of each other and with the spin axis of the body as a whole (*8*). The average visible and near-infrared colors of both lobes are indistinguishable (*9*). Near-infrared spectral absorptions on both lobes indicate the presence of methanol ice—a common, relatively thermally stable component (for an ice) of cometary bodies and extrasolar protoplanetary disks (*10*). The very red optical colors of both lobes are similar to other CCKB objects (*9*), and consistent with space weathering of simple ices to produce organic compounds, although other sources of reddening are also possible [e.g., iron and sulfur compounds (*9*)]. LL and SL both appear to be intact, or at least little disturbed, with no obvious morphological signs of a violent or energetic merger (*7, 8*).

We examine the implications of these findings for the planetesimal formation process within the Kuiper belt, which might be broadly applicable throughout the primordial Solar System. We focus on binary formation in the outer Solar System, which appears to have been common in the Kuiper belt, based on the abundance of binaries detected there in telescopic surveys (*11, 12*). A related issue is the formation of the Kuiper belt itself, its dynamical components—including the CCKB subpopulation—and the relationship between Kuiper belt objects and short period comets (*4*). Many cometary nuclei are bilobate, but because cometary surfaces and shapes have been strongly affected by solar heating—causing sublimation, mass loss and splitting— and the disruptive effects of close planetary encounters, it is not clear whether comets' bilobate shapes are a primordial characteristic or acquired during later evolution (*13-16*).

**The cold classical Kuiper belt**

Most of the bodies in the Kuiper belt are hypothesized to have been scattered and dynamically emplaced as Neptune slowly migrated outward through a massive (~15-30 Earth mass, $M_{\oplus}$) planetesimal disk that extended from ~20 to 30 au, outside the (then) compact orbits of the giant planets (e.g., *4*). CCKB objects are part of the non-resonant classical Kuiper belt located farther out, today between 42 and 47 au. CCKB objects have low dynamical excitation and physical properties distinct from the rest of the belt, so are thought to have accreted *in situ* or in nearby orbits (*17-19*). The surface density of planetesimals that built the CCKB objects, in a disk that must have extended well beyond 30 AU, was insufficient for Neptune to continue its migration past that point (*4, 20*).

The gravitational instability (GI) accretion mechanism posits that locally concentrated, gravitationally bound clouds of small (mm to dm) solid particles (the latter termed "pebbles") form in either the thick midplane of the protosolar nebular disk or in over-dense regions generated by a collective aerodynamic phenomenon called the streaming instability (*21, 22*). These concentrations then collapse directly into objects tens to hundreds of km in diameter, on time scales $\lesssim 10^3$ yr in the outer Solar System (*21-24*). GI following the streaming instability has been shown to be viable in laminar, low-viscosity (called low-$\alpha$) disks, including those with a low overall surface mass density appropriate to the CCKB (*25, 26*). Such GI models predict planetesimal formation times, velocity distributions, collisional evolution, obliquities, and binary characteristics that differ from alternative hierarchical coagulation (HC) models, in which successive two-body collisions lead to the gradual accretion of larger and larger objects (e.g., *27, 28*; see Supplementary Text.)

## Dynamic characteristics from shape and rotation

The shapes of the large and small lobes are approximately ellipsoidal, 20.6 × 19.9 × 9.4 km and 15.4 × 13.8 × 9.8 km, respectively, with a combined equivalent spherical diameter of 18.3 km (*8*). The gravitational acceleration $g$ that would be produced by the equivalent sphere is 0.0013 × ($\rho$ /500 kg m$^{-3}$) m s$^{-2}$ and the equivalent escape speed is 4.8 × ($\rho$/500 kg m$^{-3}$)$^{1/2}$ m s$^{-1}$, where $\rho$ is the bulk density. We adopt 500 kg m$^{-3}$ as our fiducial density, based on the median densities of cometary nuclei, including 67P/Churyumov-Gerasimenko, whose density has been precisely measured as 527 ± 7 kg m$^{-3}$ (*29*, their table 1) (*30*). The effective surface gravity (including the effects of rotation) across the surface of Arrokoth is shown in Fig. 1A.

If the two lobes are of equal density, the center of mass of Arrokoth is within the body of LL, and the separation between the centers of mass of the two lobes is 17.2 km [(*8*), their figure 2]. The spin-synchronous orbit period of two barely touching lobes, behaving as gravitational point masses, is $12.1 \times (500 \text{ kg m}^{-3}/\rho)^{1/2}$ hr. If Arrokoth formed as a binary pair that spiraled inwards (see below), then Arrokoth's spin must have slowed from this more rapid rotation to its observed 15.92 hour period, unless Arrokoth is substantially less dense than 500 kg m$^{-3}$. The observed rotation period of 15.92 hr would match the spin-synchronous orbit period for two barely touching, equal density lobes if $\rho \approx 290$ kg m$^{-3}$ (*8*). Explicitly treating the lobes as ellipsoids increases their mutual gravitational attraction and lowers the limiting density to $\approx 250$ kg m$^{-3}$.

The range of tensile and compressive strengths plausible for porous, structurally comet-like bodies (*29*, *31*) also broadens the permissible density range. Figure 2 shows the gravitationally induced stress at the neck, either compressional or tensile, as a function of the assumed bulk density. The contact area between the two lobes is approximately 23 km$^2$ (*8*). We calculated the gravitational attraction between the lobes from the external gravitational potential of a homogeneous triaxial ellipsoid (*32*), integrated over the mass distribution of the other ellipsoid. For the observed lobe principal axes, the attraction increases by 12.5% over a point (or spherical) approximation. A bulk density greater than ~250 kg m$^{-3}$ would imply that the neck is in compression, but even for the highest comet cohesion of 10 kPa (*33*, *34*), the density must remain under 1250 kg m$^{-3}$ or the compressive strength (which is related to the cohesion) would be exceeded and the neck region collapse under Arrokoth's self-gravity. For a more plausible, nominal bulk cometary tensile strength of 100 Pa and cohesion of 1 kPa (implying a frictional, bulk compressive strength of ~3 kPa) (*34*), the bulk density of Arrokoth must lie in the range ~250–500 kg m$^{-3}$ to explain the lack of observed faulting or distortion of the neck region (Fig. 2).

Arrokoth must possess some internal strength, otherwise it would collapse to a more spherical shape. The surface slopes with respect to the local gravity vector (Fig. 1B) are generally less than the angle of repose (maximum slope) for loose, granular material (~30–40°; *35*), so the overall shapes of LL and SL can be maintained by frictional strength alone. Near the neck, these slopes sometimes exceed 35-40°, so these over-steepened surfaces must be held together by finite cohesion (for $\rho \gtrsim 500$ kg m$^{-3}$). The minimum cohesion $c$ necessary to stabilize an inclined layer of thickness $h$ is given by

$$c/\rho g h \ = \ \tan\theta - \tan\phi \ \ \ \ , \tag{1}$$

where $\theta$ is the local slope and $\phi$ is the internal friction angle. An over-steepened thickness $h \sim 1$-2 km (Fig. 1B), $\rho = 500$ kg m$^{-3}$, $g = 10^{-3}$ m s$^{-2}$, $\theta \sim 40$-45°, and a geologically typical $\phi \approx 30°$, implies $c \sim 100$–400 Pa (*36*). This minimum strength is very low by terrestrial standards, but similar to the gravitational stresses in other low-gravity, small body environments and compatible with the interparticle forces in granular materials (electrostatic, van der Waals, etc.) (*37, 38*).

The distribution of gravitational slopes may provide additional constraints on the bulk density of small Solar System bodies (*39*). If an object possesses a sufficiently mobile regolith (surface fragmental layer), i.e., one able to overcome its intrinsic cohesion, then the surface of the body may gradually erode and/or adjust (e.g., due to impact-induced seismicity) to a state of maximum topographic stability and lowest internal stress (*39*). The distribution of slopes can therefore be related to the bulk density (subject to the aforementioned caveats). For Arrokoth's shape, there is a broad minimum in gravitational slope between bulk densities of ~200 and 300 kg m$^{-3}$ (Fig. 3), lending additional support to the inference that the density of Arrokoth may be <500 kg m$^{-3}$. If so, Arrokoth would have to be a highly porous body, given its inferred composition (*9*). Conversely, the surface of Arrokoth is only lightly cratered, so the generation of regolith and surface mobility may be inefficient (or only locally efficient, e.g., on sub-km scales, corresponding to the small-scale pitting observed (*8*)). No other Kuiper belt objects (KBOs) or cometary nuclei have confirmed densities this low, although such values have been suggested in some cases (*29*).

## Merger speed constraints

LL and SL must have merged at a very low velocity (*7*, *8*). Previous numerical simulations of collisions of km-scale (i.e., comet-like) porous icy aggregates (*15*, *33*) imply that, when extrapolated to bodies the size of LL and SL, closing velocities no greater than their mutual escape speed (several m s$^{-1}$ or less) and an oblique strike are most likely necessary to preserve the shape of a contact binary with a narrow neck. CCKBOs, even with their low-*e*, low-*i* orbits, currently have a median mutual impact speed of ~300 m/s, some 100 times higher (*40*). Thus, heliocentric impacts between bodies similar to LL and SL could not have formed Arrokoth (*7*). However, we must consider the impact velocities that would have prevailed during the early Solar System.

Arrokoth is an order of magnitude larger in size than typical comets (*8*). Therefore we performed a series of numerical experiments, modeling the collisions of bodies of the appropriate scale, density, and strength characteristics, using a soft-sphere version of the PKDGRAV N-body code (*41*, *42*) to constrain Arrokoth's formation. This code uses a discrete element method to model the collisions of granular aggregates at slow speeds and low energies, incorporating interparticle cohesion and frictional contact forces (*43*). We focus on velocities near the escape speed from the binary, which we model as two spheres for simplicity and to focus on mechanical outcomes such as the extent of distortion or disruption. The results are shown in Fig. 4.

Oblique impacts at 10 m/s, much greater than escape speed, do not lead to mergers (Fig. 4A, Movie 1), but instead shear or slice off sections of one or both bodies. The collision or merger speed of LL and SL falling from infinity (assuming an initial velocity $v_\infty = 0$) would have been $\approx 3.5 \times (\rho/500$ kg m$^{-3})^{1/2}$ m s$^{-1}$. Even oblique impacts at 5 m s$^{-1}$, slightly higher than the escape speed of 4.3 m s$^{-1}$ for the spheres in the simulations, lead to distortion and merging incompatible with Arrokoth's shape (Fig. 4B, Movie 2).

These results are essentially insensitive to the impact angle assumed. Varying the impact angle from 45° to 85° (measured with respect to the vertical at the impact point) for 10 m s$^{-1}$ collisions changes the amount of damage at the contact regions and the extent of planar shearing, but in all cases the two bodies remain unbound. At 5 m s$^{-1}$, the 45° simulation (Fig.4B) is the only one that produced a final configuration remotely resembling the present-day Arrokoth. In this case we are left with a mostly intact larger lobe, but a lopsided smaller lobe and a neck that is much thicker than observed today (Fig. 4B). At 65° and 5 m/s there is again substantial damage to the smaller lobe. At higher angles to the vertical, collisions are grazing, and as the system is initially (marginally) unbound, the simulation ends prior to the ultimate outcome (escape or re-collision).

Only at much lower collision velocities, substantially less than the mutual escape speed, and at an oblique angle, do the outcomes of our simulations begin to resemble Arrokoth (Fig. 4C, Movie 3). Movie 4 shows the maximum surface accelerations experienced by particles in the simulation shown in Fig. 4C. The disruption induced in this gentle merger (the normal velocity component is 0.5 m/s) is confined to the neck region, and more severely affects the smaller lobe portion of the neck. For a bulk density of 500 kg m$^{-3}$, varying the interparticle cohesion values over a plausible range [100 Pa to 10 kPa (*34*)] likely has only a modest effect on the gentle merger outcomes—the lobes remain should remain intact. Our numerical models show that, for Arrokoth, the merger speed of LL and SL was likely sufficiently slow that the two bodies were already gravitationally bound to each other prior to the collision. We estimate an upper limit on the vertical closing speeds of 4 m s$^{-1}$.

By way of comparison, in hierarchical collisional accretion (noted earlier), merger speeds scale with the escape speeds of the largest accreting bodies (*44*). For the cold classical region, encounter speeds could have been low initially, but would have well exceeded the above limit as the planetesimal population evolved. Our numerical models thus show it is unlikely that Arrokoth's shape could be the result of a merger of two independent heliocentric planetesimals, at any non-trivial level of dynamical excitement of the parent planetesimal swarm, unless the two lobes were much stronger (more structurally cohesive) than is usually assumed for comet-like bodies.

More generally, bilobate shapes can also be formed in catastrophic or sub-catastrophic collisions. In this scenario a contact binary results from the merger of collapsing adjacent ejecta streams following a high-speed catastrophic disruption of a parent body (*15, 45*). Such a scenario may possibly lead to a Arrokoth-like shape, if the two components first collapsed into separate bodies that then slowly came into contact. This scenario would erase any record of the precursor bodies, and in principle also permits formation of Arrokoth later in Solar System history. However, the bilobate shapes formed by these models do not resemble Arrokoth, as the lobes are not flattened and the merged components are unaligned and/or highly distorted (*15, 45*). The CCKB has not experienced strong collisional evolution (*4, 5*), making disruption of large parent bodies rare (*40*). We conclude that Arrokoth's shape and appearance are more likely the result of the low-velocity merger of two bodies that were already gravitationally bound.

**Binary formation scenarios**

We now consider how the gravitationally bound binary formed prior to the merger. Mechanisms that have been proposed for the formation of close binaries among the small asteroids may be relevant to this process. These mechanisms are not primordial in nature, but involve later collisions, or spin-up and rotational fission or mass shedding at the equator due to asymmetric solar radiation forces (*46*). Given the very low crater density on Arrokoth (*8*) and its distance from the Sun, however, these do not appear to be promising explanations (though we return to these points below). Also, secondary satellites produced from these processes are, generally, much smaller than the primary (*46*), unlike the similar sizes of SL and LL.

The prevalence of binaries in the Kuiper belt, and especially among the CCKB objects (*11*, *12*, *47*), has prompted theoretical examination of possible binary formation mechanisms specific to the Kuiper belt (e.g., *48*, *49*, *50*). Most of these mechanisms operate at Hill radius ($R_{Hill}$), the spatial limit of a body's gravitational influence in solar orbit (~4 x $10^4$ km for Arrokoth) (*51*). For example, it has been proposed that binary KBOs could form from the chance interaction of two KBOs within the Hill sphere of a third body, leaving the two permanently bound, or that dynamical friction (multiple gravitational energy and momentum exchanges) with a large number of smaller heliocentric particles could allow two passing KBOs to become bound (*52*). These mechanisms rely on energy exchanges or dissipation and thus are most effective when KBO encounter speeds are low, within an order of magnitude of the Hill speed (~2-3 cm s$^{-1}$ for Arrokoth) (*51*). Such low encounter speeds favor binding in the outer regions of the Hill sphere for retrograde orbits, or about half that distance for prograde orbits (*53*). These mechanisms thus geometrically favor the production of retrograde binaries, sometimes strongly so (*54*), but observations show that prograde binaries are more common than retrograde (*55*). Such chance encounters of KBOs would produce some binaries with different color characteristics within each pair. This also disagrees with observations, which show that KBO binaries all share the same colors (*12*, *47*, *56*). We therefore discard these models in favor of a binary formation mechanism that produces both bodies from a compositionally uniform region of the protosolar nebula.

An alternative binary formation mechanism posits swarms of locally concentrated solids in the protoplanetary disk that collapse under self-gravity. The swarms of particles could form as concentrations produced by the streaming instability (SI), in which the drag felt by solid particles orbiting in a gas disk leads to a back-reaction and spontaneous concentration of the particles into massive filaments and clumps (Fig. 5A), which can then gravitationally collapse (*22*, *23*, *24*, *57*). The collapse mechanics have been simulated for the formation of larger, 100-km class Kuiper belt binaries (*58*). That work simulated bound particle clumps in 3D with the PKDGRAV N-body code, including collisions and assuming perfect merging (100% sticking). Rotating particle clumps in (*58*) typically collapse to form binaries or higher multiple planetesimal systems (Fig. 5B). The mechanism produces binaries with a broad range of separations and eccentricities, depending on the initial swarm mass and angular momentum (*58*, their fig. 5). The resulting binary orbital parameters are consistent with observations of binaries in the classical Kuiper belt (*11*, *12*, *58*) including the approximately equal radius ratios of the binary components (and of Arrokoth) (*47*, *59*). We also expect such binaries to have matching component colors, as they formed from the same material.

The angular momentum vector orientations of collapsing particle clouds have been estimated (*60*). That work fully simulated vertically stratified 3D hydrodynamical SI (following

*61, 62*), identified gravitationally bound clumps of solid particles (Fig. 5A), and determined the obliquity of the corresponding angular momentum vectors. The total angular momentum stored in a particle clump typically exceeds the maximum possible for a compact object of the same solid mass (set by the rotation speed at which it would break up) (*60*). Therefore as a particle clump contracts and speeds up, it must either shed mass and angular momentum, or form binary or higher-multiple systems. The simulations in (*60*) did not reach binary formation [unlike (*58*)], due to computational resolution limits. The resulting angular momentum vectors of the gravitationally bound clumps, however, span a range from prograde to retrograde, with a strong preference for prograde over retrograde rotation (*60*). This is consistent with observations of KBO binaries (*55*), even with the broad range of obliquity produced by the inherently stochastic, turbulent nature of clump collapse process (*58*). In addition, previous results (*58, 47*) indicate that a fraction of the non-binary-forming solids in a contracting clump are expelled from the clump into the general nebular population (Fig. 5B). These accretional products and any surviving, unaccreted pebbles are then available for further cycles of concentration due to the SI (or other mechanisms).

Possible mechanisms to produce particle density enhancements in the outer protosolar nebula, acting individually or together, and which could have led to GI, include the SI, photoevaporation, pressure bumps or traps, and volatile-ice lines (*57, 63*, Supplementary Text). Clumping due to the SI in particular is consistent with the mass function of the CCKB and Arrokoth specifically (Supplementary Text).

If Arrokoth initially formed as a co-orbiting binary, a subsequent step of orbit contraction is required in which angular momentum is lost, ultimately resulting in a binary merger. For a gravitationally collapsing pebble cloud (*58*), such a merger may happen directly if the angular momentum density is low enough. In a higher-mass cloud, or in one with higher angular momentum density, a smaller-mass binary, co-orbiting or not, may be expelled from the collapsing cloud (Fig. 5B). The shape and alignment of Arrokoth's lobes constrain the nature of any orbital contraction, discussed next.

**Lobe shape and alignment**

The global, contact binary shape of Arrokoth (*7*, *8*, Fig. 1) is reminiscent of co-orbiting Roche ellipsoids in close contact. Roche ellipsoids are the equilibrium shapes of rotating homogeneous fluid masses distorted by the tidal action of a nearby more massive body (*64*, Supplementary Text). However, the flattened shapes of the observed lobes (axis ratio ~1/2 for LL and ~2/3 for SL) do not match a Roche ellipsoid, because the less massive lobe should be more oblate than the more massive one, even when considering higher-order gravity terms and internal friction (e.g., *65*). The present-day shape of Arrokoth does not conform to an equipotential surface at any uniform density or rotation rate (*8*).

The generally ellipsoidal to lenticular shapes of Arrokoth's two lobes, and their general smoothness at scales resolved by the available images (*7*, *8*), nevertheless resemble equilibrium figures, perhaps obtained in the past. It is possible that the flattened shapes of both lobes were acquired as they rapidly accreted in a pebble cloud undergoing gravitational collapse, as described above. The spin rates necessary to reach the observed flattened shapes would have been higher than Arrokoth's spin today, but not by a large margin. For low density (250 kg m$^{-3}$), strengthless oblate bodies (Maclaurin spheroids, *64*), the rotation periods of LL and SL would need to have been ~12 and 14 hr, respectively [these values scale as $\rho^{-1/2}$ (*64*)], compared with the current rotation rate of 15.9 hr. The process(es) that collapsed the co-orbiting binary, as described below, could have potentially slowed the spin the individual lobes by this amount.

Regardless of the origin of the shapes of the two lobes (Supplemental Text), the close alignment of their principal axes (Fig. 6) (*7*, *8*) is unlikely to be due to chance alone. The short axes (which we designate as *c*-axes) of LL and SL are closely aligned, to within 5°, a value set by systematic uncertainties in the shape models (*8*). The long *a*- and intermediate *b*-axes are aligned as well, but the *a*- and *b*-axes of LL are similar in length (20.6±0.5 and 19.9±0.5 km respectively, 1σ uncertainties), so the alignment of SL's *a*-axis with the long axis of the body as a whole (also within 5°) is more meaningful (Fig. 6A). These angles are small enough to be considered in sequence. The *c*-axis of one lobe must be lie within a cone of half-angle 5° with respect to the *c*-axis of the other ([1- cos(5°)] = 0.0038); with that orientation fixed, the *a*-axis of SL must lie within 5° of the long axis of the body as a whole (10/180 = 0.056). The joint probability of both aligning due to chance is ~0.0038 × 0.056 = 2 × 10$^{-4}$.

We infer that before their final merger, the LL and SL lobes were already aligned. That would be consistent with tidal evolution of a close binary, as alignment reduces the total energy of the system. Full spin-orbit synchronism (tidal locking) is not required, however. Two irregular bodies rotating asynchronously while their mutual semimajor axis slowly shrank (by any mechanism) would necessarily first contact each other along their long axes, perhaps repeatedly, if the orbits were circular (the same outcome is likely but not guaranteed for elliptical orbits). Ultimately, mechanical dissipation of rotational kinetic energy while in contact would cause their long axes to come to rest in alignment (or nearly so; Supplementary Text).

Regardless whether the *a*- and *b*-axes of Arrokoth were aligned prior to the merger, the *c*-axes must have been. Merger, even a slow merger, from an arbitrary direction is very unlikely.

Chaotic rotation (tumbling) of either lobe, owing to an eccentric orbit pre-merger (*66, 67*), highly unlikely to produce this alignment. The LL and SL lobe spin poles, and their mutual orbit normal vector, were most likely close to co-aligned prior to merger, consistent with mutual tidal dissipation. This may also be a common (though not exclusive) outcome of binary formation in a gravitationally unstable pebble cloud (*58*), with its shared angular momentum. Although previous work focused on wide binaries (*58*), it is possible to form binaries with a range of orbital separations, including those much closer to contact, or already contacting, within the collapsing cloud.

**Binary merger mechanisms**

The GI formation mechanism produces a high fraction of binary Kuiper belt objects, but as discussed above, the resulting angular momentum of each binary may have been greater, perhaps much greater, than the current Arrokoth system. Here we consider several, non-mutually exclusive mechanisms that might drain angular momentum from the system over all or part of its 4.5-billion-year lifetime.

*Kozai-Lidov cycling.*

In a system with three (or more) bodies with differing orbital inclinations, the Kozai-Lidov effect (*68*) causes oscillations of the orbits eccentricity and inclination. We focus on the Sun as the third body, and the Kozai-Lidov cycles of the orbits of LL and SL about each other. In this case the angular momentum component of the binary perpendicular to the heliocentric orbital plane is conserved. On time scales much longer than Arrokoth's 298-yr heliocentric orbital period (following [*68, 69*], $\sim 10^5$ yr $\times$ ($a$/1000 km)$^{-3/2}$, where $a$ is the assumed LL-SL semimajor axis), highly inclined, near-circular orbits can transition to and from low inclination, highly eccentric orbits. During periods of high eccentricity, the binary objects pass closer to one another, and so have stronger tidal interactions (*70, 71*). If the eccentricity becomes sufficiently high, the objects could undergo grazing collisions that would substantially alter the balance between orbital and rotational angular momentum, and efficiently dissipate kinetic energy. High eccentricity phases also cause objects to spend most of the time near their maximum separation (apoapse), where they are more susceptible to perturbation by unbound bodies passing through the system.

Solar tides are weak in the Kuiper belt, and the Kozai-Lidov cycles occur slowly. Solar tides are not important except for comparatively wide binaries, because the tides due to non-spherical shapes can dominate the dynamics of closer binaries. For Arrokoth in particular, solar perturbations would only dominate at binary semimajor axes *a* >1000 km (~100 LL radii) (*72*). Also, if Kozai-Lidov oscillations had affected Arrokoth, we expect the merged body (in most cases) would have a lower obliquity than the observed 99°, because the tidal interactions or collisions at high orbital eccentricity would have tended to lock in the low inclinations that correspond to the highest eccentricities (*69-71*).

An alternative possibility is that Arrokoth was once a triple system, and that the third body was in an inclined orbit with respect to the then LL-SL binary. For suitable orbital parameters, this third body could have driven Kozai-Lidov oscillations of the inner binary. Hierarchical triple systems do exist among small [e.g., KBO 47171 Lempo (*73*)], and are a common outcome of simulations (*58*). But because there is no specific evidence of a lost third body, we do not consider this hypothesis in greater detail.

*YORP and BYORP.*

Interaction with sunlight can affect the angular momentum of small bodies in two main ways: the Yarkovsky–O'Keefe–Radzievskii–Paddack (YORP) effect, which alters the spin rate and obliquity of a single object, and binary YORP (BYORP), which changes the size and shape of a binary's orbit (*74*, *75*). Both mechanisms arise from the asymmetric scattering and thermal re-emission of sunlight from the surfaces of irregular bodies, and both can either increase or decrease the angular momentum of the system and alter its vector direction (*74*, *75*).

BYORP can in principle drain the angular momentum of a binary near-Earth asteroid, provided one or both members of the binary are spinning synchronously (*76*). BYORP requires $10^4$ to $10^5$ years to alter the orbit of a 150-m radius, synchronously rotating satellite of a 500-m radius primary, both with density 1750 kg m$^{-3}$, assuming a satellite orbital radius of 4 primary radii and a primary orbit at 1 au (*76*). Scaling that result to parameters (size, distance, density, mass ratio) appropriate to Arrokoth (*77*), we obtain a time scale of a few billion to a few tens of billon years, a span that includes the age of the Solar System. Thus the two components of Arrokoth, if initially separated by a few LL radii, could in principle be driven by BYORP radiation forces alone into a gentle merger of the type needed to account for the narrow neck connecting the two bodies, albeit late in Solar System history (*78*).

YORP accelerations (unlike BYORP) have been detected for several asteroids (Table S1). Asteroid (1862) Apollo, at 1.5 km across and orbiting at 1.5 au, exhibits the largest, measured YORP coefficient (*Y*), with an estimated YORP spin doubling time scale of ~$5 \times 10^5$ years for an assumed density of 2500 kg m$^{-3}$ (Table S1). Accounting for the larger size of Arrokoth, its greater distance from the Sun, and its much lower density (*77*) lengthens this time scale to 7.5 $\times 10^9$ years. This exceeds the age of the Solar System, and adopting any of the other (lower) *Y* values in Table S1 would imply even longer time scales for Arrokoth, by up to two orders of magnitude. However, a more rapidly-spinning contact binary in the Kuiper belt, such as might be produced by BYORP, could plausibly be slowed by subsequent YORP torques over the age of the Solar System (e.g., from a 12-hr period to its present 15.9 hr).

The shape and surface properties of Arrokoth and any obliquity variations could change the strength and sign of the YORP torques (*79-82*). Also, both YORP torques perpendicular to the spin axis and BYORP torques perpendicular to the orbit normal vector are minimized at high obliquities and binary inclinations, respectively (*76, 81*). Arrokoth's high obliquity is therefore less likely to have changed much over the age of the Solar System due to radiation effects, even if its spin has.

*Tides.*

Tides could have contributed to Arrokoth's orbital evolution when the two lobes were close, including establishing the synchronous spin-orbit locking necessary for BYORP torques to have been effective. We calculate the spin-down time scale for SL due to tides from LL using standard methods (*83, 84*). Tidal evolution from the breakup spin limit (*85*) to its present value would take ~65-to-650 × ($a$/100 km)$^6$ Myr, assuming a circular orbit and adopting a bulk density of 500 kg m$^{-3}$ for both lobes and tidal dissipation parameters for SL (*84, 86*). If the Arrokoth binary originally formed within 100-200 km (≲25 LL radii), or $a$ was driven below that limit by other processes, tides would have dominated. For $a$ < 50 km, tidal synchronization of SL's spin would have been rapid (≲1-10 Myr).

On their own, tides between LL and SL do not shed angular momentum, but redistribute it among the individually rotating lobes (including aligning their spin and orbital angular momenta). If LL were rotating more slowly than SL's mean motion, for example, tides would act to shrink the binary orbit, and at the moment of tidally-induced contact the overall rotation rate of the merged binary would jump abruptly. A more slowly rotating LL could have resulted from YORP torques, which affect individual binary components even when the components are not rotating synchronously. Tidal interactions within a hierarchical triple system could also have led to angular momentum exchange (as noted above), and loss from the system if the more distant member of the triple ultimately escapes to heliocentric orbit (*68*).

*Collisions.*

Bilobate comets, such as 67P, have led to the suggestion that mutually orbiting binaries in the Kuiper belt may have their binary orbital angular momentum altered by repeated impacts with smaller, heliocentric planetesimals, resulting in a contact binary (*87*). Impacts can both bind or unbind a binary—with the binary's orbital angular momentum executing a random walk—so binding to coalescence is only probable (though by no more than ~30%) for very close binaries (*87*). Also, the heliocentric impactor flux in the CCKB object region is estimated to be (and to have always been) low, and deficient in smaller, sub-km-scale bodies (*8, 40*), making this mechanism unlikely.

Arrokoth's low crater density (*7, 8*) also makes impacts an unlikely candidate for collapsing the pair's orbit. Only formation of the largest impact crater, informally named Maryland (*8*), could have substantially affected the angular momentum of Arrokoth. Assuming an impactor diameter of ~1 km (1/7 the diameter of Maryland), an impact speed of 300 m s$^{-1}$ [typical for Arrokoth impactors (*40*)], an impact angle of 45°, and an optimistic impact orientation (a velocity vector in Arrokoth's equatorial plane), Arrokoth's total angular momentum only changes by ~10% if $a$ was ~100 km at the time of the impact. The transfer of linear impactor momentum to binary angular momentum scales as $a^{1/2}$, so the formation of Maryland could have had a stronger effect if Arrokoth formed originally as a wide binary. All other observed impact craters are much smaller.

*Gas Drag.*

Drag may have been exerted on the binary by protosolar nebular gas. Within a collapsing pebble cloud, the mean collision time is shorter than the gas-drag stopping time (the time it takes for a pebble's linear momentum to drop by a factor of *e*) (*58*). This implies that binary formation and dynamics during GI are dominated by collisions and dynamical friction, not intracloud gas dynamics. Once the unaccreted cloud remnant disperses, however, the binary is subject to gas drag forces for as long as the gas in the protosolar nebula persists at Arrokoth's heliocentric distance (*88*).

The momentum flux (due to gas drag) imparted by an ambient gas to an orbiting binary yields a stopping time of $t_{stop} \sim \rho R/(\rho_{gas} \upsilon_{orb})$ (*89*), where $R$ is the mean radius of either the primary or secondary and $\rho_{gas}$ is the gas density, assuming a drag coefficient $C_D$ of ~1, which is appropriate to fully turbulent drag (see below). Adopting a characteristic midplane $\rho_{gas}$ at 44.2 au of $1 \times 10^{-10}$ kg m$^{-3}$ (*90*) and an initial semimajor axis for Arrokoth of 100 km yields an orbital speed $\upsilon_{orb}$ ~1 m s$^{-1}$ and stopping times of ~500 Myr for $\rho = 500$ kg m$^{-3}$ and an average $R = 7$ km (with gas drag acting on each lobe). This is much longer than any plausible lifetime for the protosolar nebula, likely no more than 10 Myr (*90, 91*), so it might seem that ambient gas had little effect on Arrokoth's later evolution.

The gas drag environment experienced by Arrokoth is, however, likely to have been more complex than that simple calculation. The nebular gas at Arrokoth's distance from the Sun would have been moving at speeds slower than the equivalent Keplerian orbit owing to the pressure gradient in the nebula (*88, 92*):

$$\Omega^2 r \;=\; \Omega_{\mathrm{K}}^2 r + \frac{1}{\rho}\frac{\partial P}{\partial r} \qquad , \qquad\qquad (2)$$

where $\Omega$ is the angular velocity of the gas, $\Omega_{\mathrm{K}}$ is the Keplerian angular velocity due to the Sun's gravity, $P$ is the gas pressure, and $r$ is the heliocentric distance. Because Arrokoth itself orbits the Sun at Keplerian speed, it will feel a headwind (at velocity $\upsilon_{\mathrm{wind}} = r(\Omega_{\mathrm{K}}-\Omega)$), which we estimate (from *90*) to be ~50 m/s, about 1% of the Keplerian speed. This gas velocity determines the drag regime at Arrokoth, irrespective of the binary's orientation, and couples to the slower velocity of the co-orbiting binary. As the Arrokoth binary orbits in this nebular wind (Fig. 7), each of its lobes will alternately feel accelerating and decelerating torques; time averaged over both the binary's mutual and heliocentric orbital periods, the difference is proportional to $\upsilon_{\mathrm{orb}}\upsilon_{\mathrm{wind}}$, resulting in a modified stopping time (time to reduce the binary's angular momentum by a factor of $e$)

$$t_{\mathrm{stop,wind}} \sim \frac{\rho R}{C_{\mathrm{D}}\rho_{\mathrm{gas}}\upsilon_{wind}} \qquad\qquad , \qquad (3)$$

where the drag coefficient $C_{\mathrm{D}}$ is now explicitly included (and the high obliquity of Arrokoth is included as well).

   The kinematic viscosity ($\eta$) of solar nebula gas, for the above midplane conditions, is ~$10^5$ m$^2$ s$^{-1}$ (*93*), which in turn implies Reynolds numbers $Re \equiv 2R\upsilon_{\mathrm{wind}}/\eta \sim 15$ for Arrokoth. This puts Arrokoth into the intermediate drag regime (*92, 94*), with corresponding $C_{\mathrm{D}}$ values of $24Re^{-0.6} \sim 5\text{-}10$ for its two, non-spherical lobes. Combined with the wind-speed dependence in the time-averaged torque, the gas-drag stopping time from Eq. (3) (a measure of the binary merger time scale) decreases by a factor of ~250-500, to ~1–2 Myr for Arrokoth. Such time scales are commensurate with the short lifetimes of protoplanetary gas disks (e.g., *96, 97*). Alternative protosolar nebula models (e.g., *88, 94, 95*) yield comparable time scales.

Headwind-coupled gas drag may therefore be the dominant mechanism that drove the merger of small Kuiper belt binaries such as Arrokoth. In the intermediate-*Re* drag regime, the merger time scales as $\rho R^{1.6}$ (*92, 94*) so smaller binaries (e.g., similar to comet 67P in scale) would have evolved to become contact binaries even more rapidly. The effects of gas drag do not cease once the contact binary forms, though the geometry of the drag interaction become more complicated. Low-inclination binaries would shrink faster than high-inclination binaries (by a factor of $\sim \pi/2$), all other things being equal, because the headwind is always edge-on to their mutual orbits. This leads us to predict that for a given distance from the Sun, the physical sizes of low-inclination contact binaries extend to larger scales than high-inclination contact binaries. There may also be a complementary excess of more distant co-orbiting binaries at high inclinations.

We adopted a specific nebular density profile (*90*) above because it is consistent with the initial compact giant planet configuration and outer planetesimal disk thought to have been present in the early Solar System (e.g., *4*). This profile was designed to represent the protoplanetary nebula at the time of planetesimal formation. It also assumes that the nebula (gas and solids) does not end abruptly at $\sim 30$ AU, but gradually declines in surface density to satisfy the constraint that Neptune's outward migration ceases at that distance (*20*). If the gas nebula was instead highly attenuated in the CCKB region, gas-drag-driven binary merger would have been ineffective. Because the characteristics of Arrokoth indicate planetesimal formation via the SI or a related collective instability, we nevertheless conclude there must have been sufficient gas and, at least locally or intermittently, sufficiently high solid/gas ratios for planetesimal forming instabilities to occur.

## Arrokoth's Story

Numerous mechanisms have been proposed to produce macroscopic bodies from small particles in the protosolar nebula. The *New Horizons* encounter with Arrokoth has allowed those mechanisms to be tested with close observation of a primitive planetesimal.

Arrokoth is a contact binary (*7*), consistent with being a primordial planetesimal (*7-9*). There is no evidence of heliocentric, high-speed collisional evolution, or any catastrophic (or even a sub-catastrophic) impact during its lifetime. Its shape is not consistent with hierarchical accretion of independent, heliocentric planetesimals, as initially slow collisions would have eventually become catastrophic. Instead, we conclude that its two lobes (LL and SL) came together at low velocity, no more than a few m/s and possibly much more slowly.

Binary formation is a theoretically predicted common outcome in protoplanetary disks when swarms of locally concentrated solids (pebble clouds) collapse under self-gravity, which plausibly explains the high fraction of binaries among cold classical KBOs (*58*). Cold classical KBO binaries exhibit a range of binary orbital separations, down to the presently observable limit (~1000 km [*47*]). Numerical modeling indicates that tighter or contact binaries could form in a collapsing pebble cloud. The prominence of bilobate shapes among the short period comets, which are derived from the scattered disk component of the Kuiper belt, suggests (but does not require) that there is a process that collapses Kuiper belt binary orbits (*87*). The alignment of the principal axes of the LL and SL lobes indicates tidal coupling between two co-orbiting bodies, prior to their final merger.

Our examination of various mechanisms to drive binary mergers in the Kuiper belt indicates the potentially dominant role of gas drag while the protosolar nebula is still present. We find this process to be effective, because in a gas nebula with a radial pressure gradient the velocity of the gas deviates from the heliocentric Keplerian velocity of the binary. The headwind that the binary feels couples to the motion of the binary pair about its own center of mass. The resulting viscous gas drag can collapse Arrokoth-scale co-orbiting binaries—as well as smaller, cometary-scale binaries—within the few-Myr lifetime of the protosolar gas nebula.

The presence of substantial nebular gas in the region of the cold classical Kuiper belt does not conflict with the low planetesimal mass density in the same region (*4*). Gas drag drift of small particles can cause large-scale depletion of the solids in the cold classical region (*92*, *94*). Enough solid mass must nevertheless have remained to build the cold classical KBOs, a population that has likely dynamically lost only a few times its present mass over Solar System history (*19*). Collective, gravitational instabilities in the presence of nebular gas can produce a planetesimal population from such a low solid mass density. Similar accretional processes may have occurred elsewhere in the early Solar System.

**Acknowledgments:** These results would not have been possible without NASA, and the efforts by the *New Horizons* search, occultation, encounter, and navigation teams to discover, locate, and precisely rendezvous with Arrokoth. We thank the reviewers for their perceptive comments and E. Asphaug and M. Jutzi for informative discussions. **Funding:** This research was supported by NASA's *New Horizons* project via contracts NASW-02008 and NAS5-97271/TaskOrder30. J.C.M. and D.C.R. were supported by NASA Solar System Workings grant NNX15AH90G, and J.C.M. by a University of Maryland Graduate School Research and Scholarship Award. J.J.K. was supported by the National Research Council of Canada. **Author contributions:** W.B.M. led the study and wrote the paper with D.C.R., J.C.M., J.T.K., D.P.H., O.M.U. and W.M.G., with inputs from S.A.S., D.N., T.R.L., K.N.S., and H.A.W.; J.C.M. and D.C.R. performed the PKDGRAV collisional simulations; S.B.P. led the development of the shape model; X.M. calculated the ellipsoidal gravity; W.B.M., J.T.K., J.C.M. and D.C.R. produced the figures; M.W.B., J.R.S., M.R.S.C.M.L., J.J.K., J.W.P. and A.H.P. contributed to the discussion; D.P.C., H.A.E., G.R.G. and J.M.M. lead the *New Horizons* Science Theme Teams, and S.A.S., H.A.W., J.W.P., C.B.O., K.N.S., A.J.V. and L.A.Y. are lead scientists of the *New Horizons* project. The entire *New Horizons* Science Team contributed to the success of the Arrokoth encounter.

**Competing interests:** The authors declare no competing interests. **Data and materials availability:** Executable PKDGRAV code, along with input and output files for the results presented in this paper, are available at DOI 10.6084/m9.figshare.11653167.

**Supplementary Materials:**

Team Members and Affiliations

Materials and Methods

Supplementary Text

Table S1

References (*101-118*)

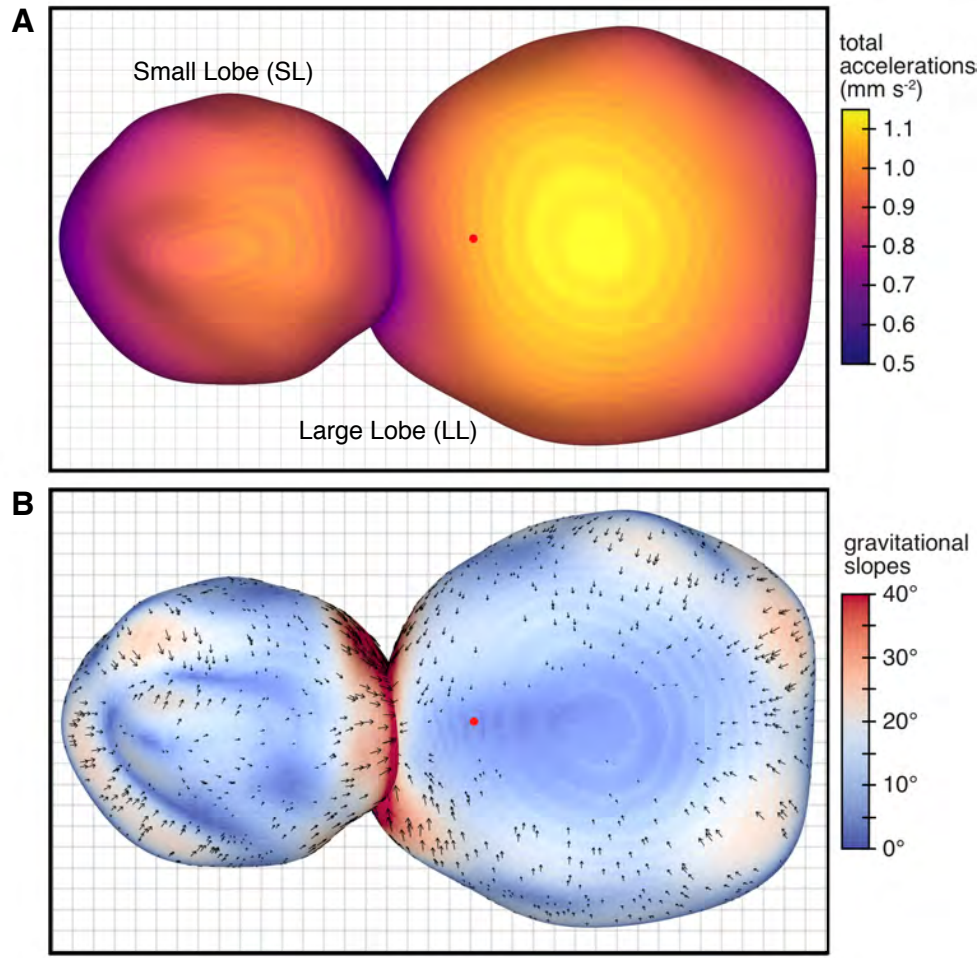

**Fig. 1. Dynamic geophysical environment at the surface of Arrokoth is determined by its gravity and rotation.** (**A**) Effective surface gravity in Arrokoth's rotating frame, overlain on the shape model (*8*). (**B**) Gravitational slope, i.e., the difference between the local effective gravity vector and the surface normal to the global shape model. Arrows indicate the tilt of the local gravitational slope; the steepest slopes occur in or near the neck region (cf. (*8*), their figure S1). In both (**A**) and (**B**) a uniform density of 500 kg m$^{-3}$ is assumed for both lobes; red dot indicates center of mass/rotation axis. The background grid is in 1 km intervals.

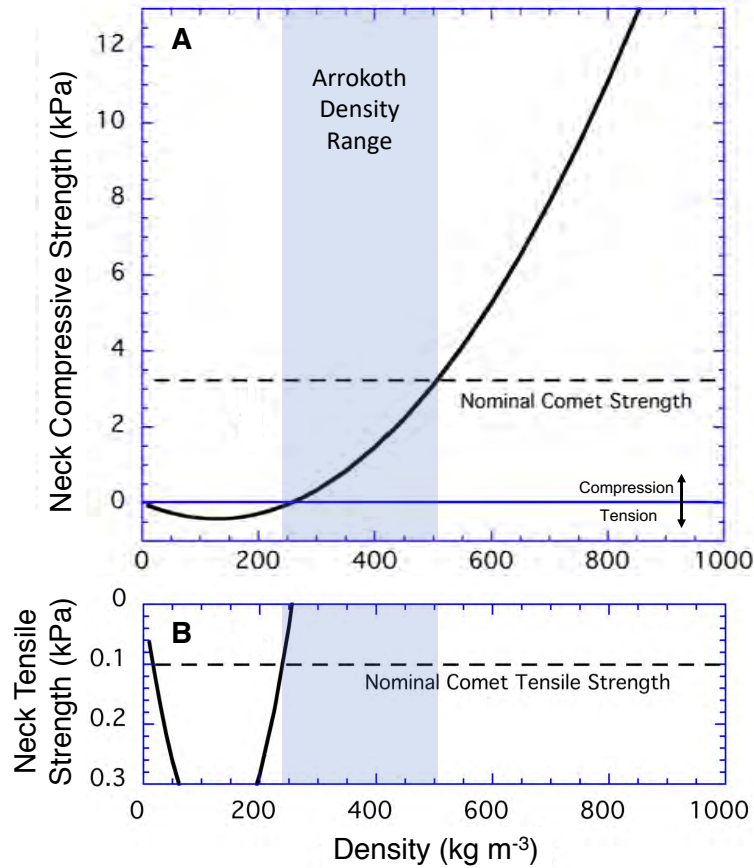

**Fig. 2. The compressive or tensile stress supported at Arrokoth's neck depends on the body's density.** (**A**) The solid blue line separates the unconfined compression and tension regimes. For bulk densities $\rho \gtrsim 250$ kg m$^{-3}$, the neck is in compression. For a nominal cometary cohesion of 1 kPa (dashed black line) and internal friction angle of 30° (*33*), the upper limit density of Arrokoth is ~500 kg m$^{-3}$; otherwise the neck region would collapse. Greater strengths are compatible with greater bulk densities. For $\rho \lesssim 250$ kg m$^{-3}$, the neck is in tension (shown on an expanded scale in **B**). For a nominal cometary tensile strength of 100 Pa (*33*) (dashed black line), the lower limit density of Arrokoth remains close to 250 kg m$^{-3}$. Much lower densities ($\lesssim$ 50 kg m$^{-3}$), for which the forces between the lobes vanish, are not considered physical. Strength estimates scale inversely with the assumed contact area (we adopted 23 km$^2$ [*8*]).

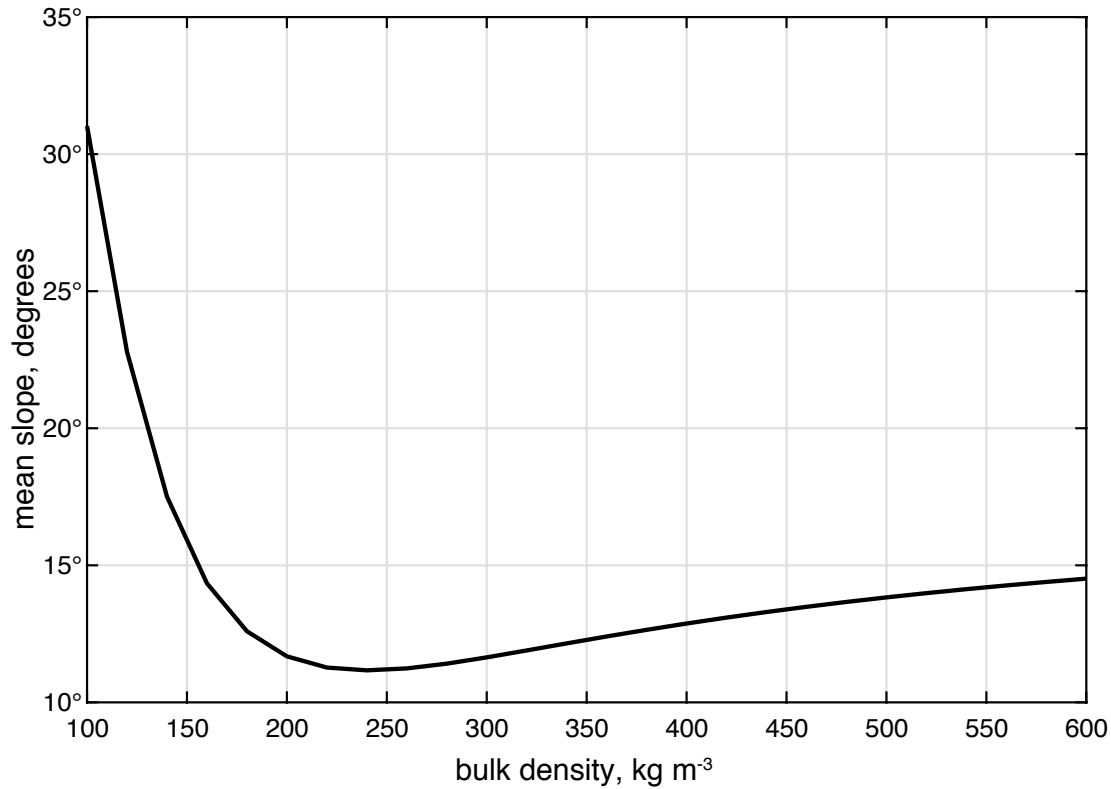

**Fig. 3. The mean gravitational slope of Arrokoth as a function of assumed bulk density.**

The minimum mean slope occurs for a bulk density of ~240 kg m$^{-3}$ (cf. Fig. 1B, which assumes $\rho$ = 500 kg m$^{-3}$). If Arrokoth's topography behaves similarly to that of asteroids and cometary nuclei (*39*), this may be the approximate density of Arrokoth. The minimum is quite broad, however, which is consistent with a range of densities considered appropriate to cometary nuclei (*29*).

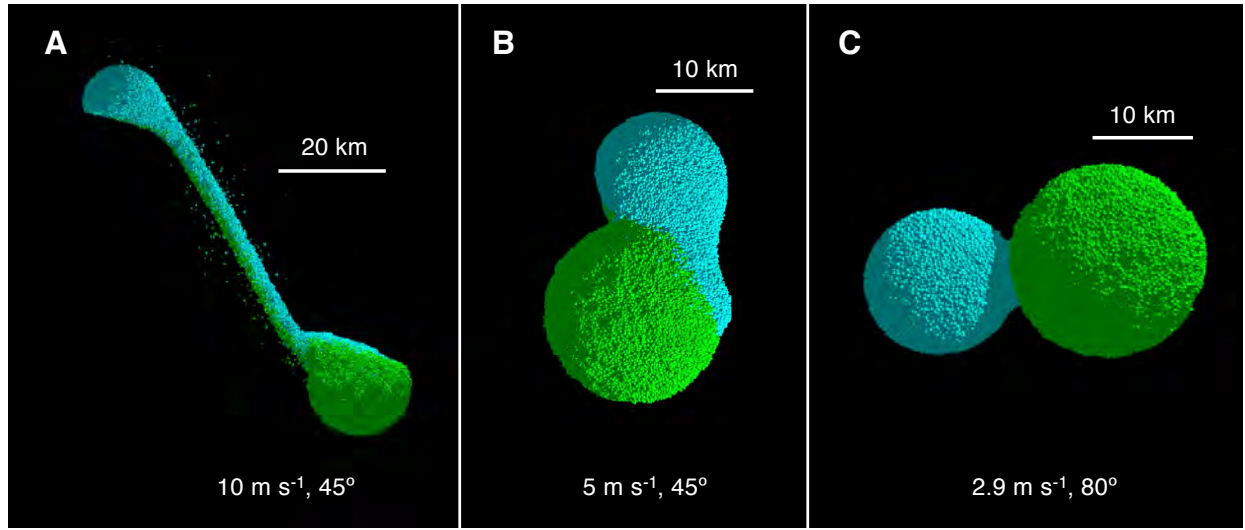

**Fig. 4. Numerical N-body calculations of collisions between spherical bodies of the scale and approximate mass ratio of the LL and SL lobes of Arrokoth**. The larger lobe (LL) is represented by green and the smaller lobe (SL) by blue particles, respectively. A bulk density of 500 kg m$^{-3}$ is assumed for both bodies. (**A**) At a collision speed of 10 m s$^{-1}$ and a moderately oblique angle, the impact severely disrupts both bodies, leaving a long bridge of material stretched between them. As the simulation progresses, this connection breaks as Thule moves farther from LL and ultimately escapes. Movie 1 shows an animated version of this panel. (**B**) At 5 m s$^{-1}$ and for the same impact angle of 45°, the impact creates a contact binary, but with an asymmetric, thick neck and a lopsided Thule. Movie 2 shows an animated version. (**C**) At 2.9 m s$^{-1}$ and an oblique impact angle of 80°, both lobes remain intact, and the contact area between them forms a well-defined, narrow neck. Movie 3 shows an animated version. Interparticle friction between the particles is assumed in all cases; in **A** and **B** the interparticle cohesion is 1 kPa (a value thought typical for comet-like bodies; see text) and zero cohesion is assumed in **C**. No initial spin is assumed in **A** and **B**, whereas the lobes in **C** are set to rotate synchronously before collision.

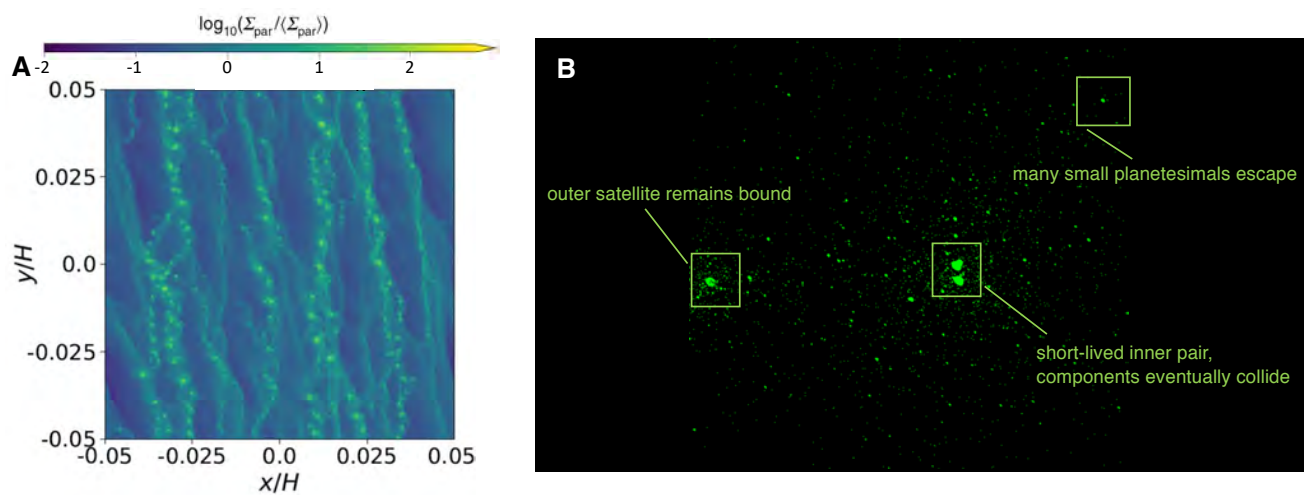

**Fig. 5. Possible initial stages in the formation of a contact binary in the Kuiper belt, illustrated by numerical models.** (**A**) Overdense particle concentrations in the protosolar nebula self-amplify by the streaming instability, which then leads to gravitational instability and collapse to finer scale knots. Snapshot from a numerical simulation in (*60*) illustrates vertically integrated particle density, $\Sigma_{par}$, viewed perpendicular to the nebular midplane, relative to the initially uniform surface density, $\langle \Sigma_{par} \rangle$; lighter colors mean greater particle density, $H$ is nebular scale height, and $0.02H$ is the initial particle scale height (figure adapted from (*60*); © reproduced with permission). (**B**) Outcomes of an example collapsing, gravitationally unstable particle cloud, from N-body simulations in (*58*). Arrokoth may have formed as a binary planetesimal in such a collapsing particle cloud, either as a contact, or more likely, a co-orbiting binary.

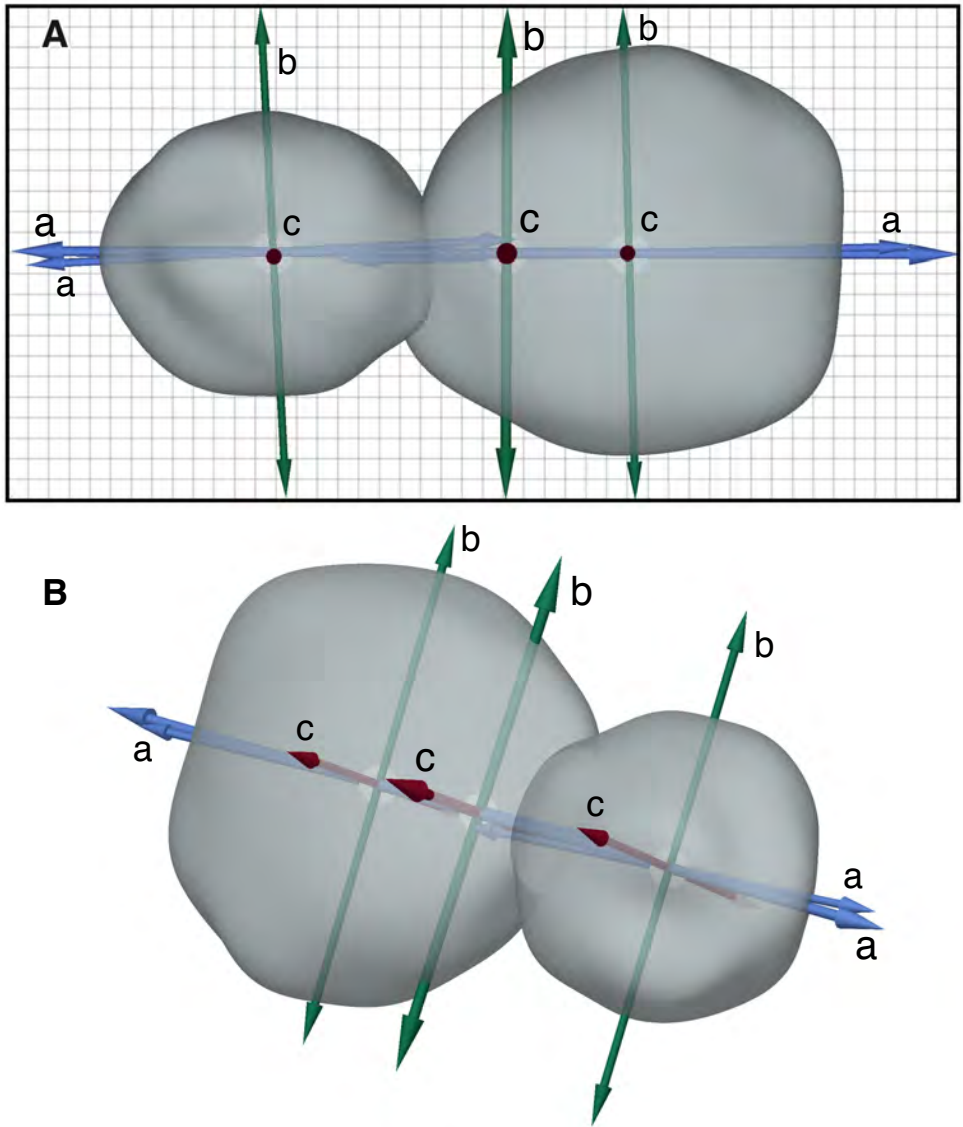

**Fig. 6.  The inertial axes of Arrokoth and its two lobes are aligned.** (**A**) Viewed down the spin axis, arrrows indicate the maximum (c, or red), intermediate (b, or green), and minimum (a, or blue) principal axes of inertia for each lobe (thin vectors), and the body as a whole (thick vectors). Vectors originate from the center of mass of each component. Background grid is in 1-km intervals. (**B**) Oblique view of the same, matching the geometry of the CA06 image (*8*). Alignment of the maximum principal axes of inertia, and of SL's minimum principal axis of inertia with that of Arrokoth as a whole, is unlikely to be due to chance alone (see text).

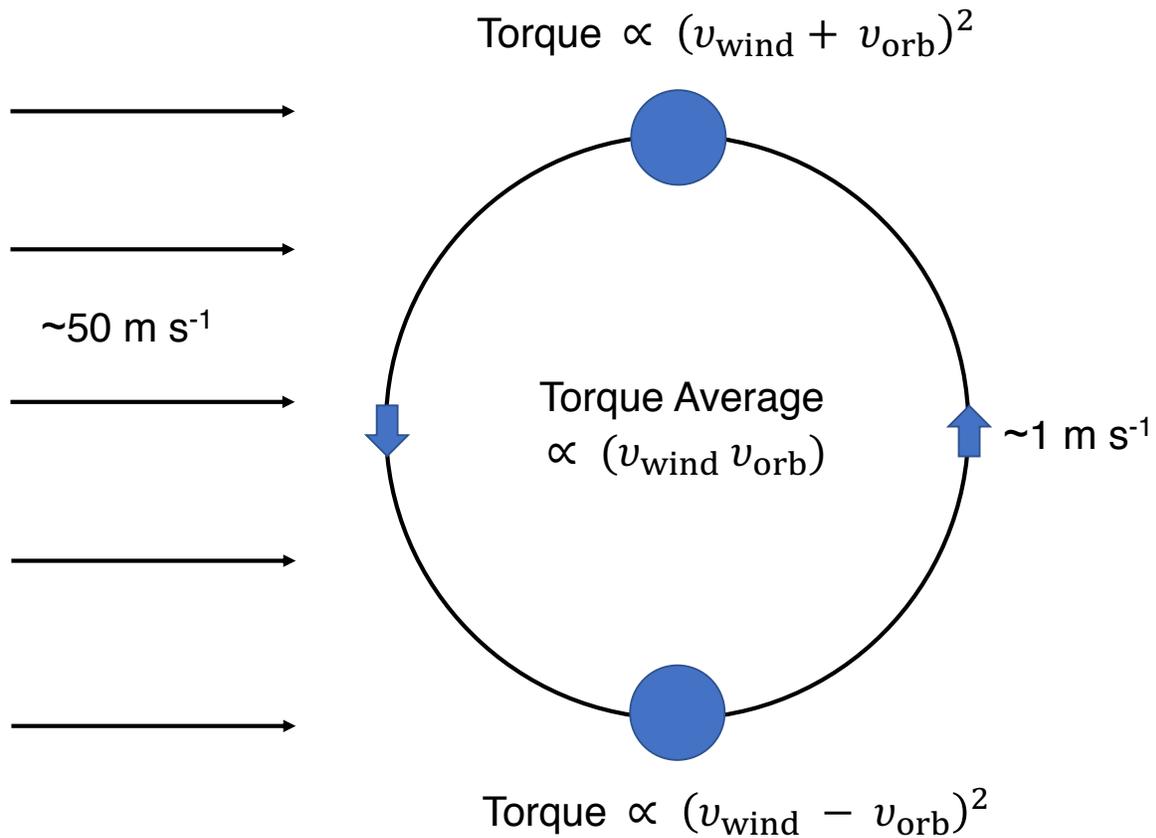

**Fig. 7. Illustration of the protosolar nebula headwind interacting with a co-orbiting equal mass binary.** The averaged torque is proportional to the product of the lobe orbital velocity and the differential velocity between the nebular gas and the binary's center-of-mass about the Sun.

**Movie 1.**

**Animated version of Fig 4A.** Merger of spherical components at 10 m s$^{-1}$ and impact angle 45°, using a rubble-pile model with 198,010 particles total. Material parameters correspond to rough surfaces with a friction angle of ~40° and a cohesion of 1 kPa. For this model there was no initial component rotation. Particle color indicates body of origin – green particles are from the large lobe (representing LL), while blue particles are from the small lobe (representing SL).

**Movie 2.**

**Animated version of Fig 4B.** Merger of spherical components at 5 m s$^{-1}$ and impact angle 45° degrees using a rubble-pile model. Particle size and density and material parameters are identical to the simulations in Movie 1. For this model there was no initial component rotation. Particle color indicates body of origin.

**Movie 3.**

**Animated version of Fig 4C.** Merger of spherical components at 2.9 m s$^{-1}$ and impact angle 80° using a rubble-pile model. Particle size and density and material parameters are identical to the simulations in Movies 1 and 2. For this model, each component has an initial spin period of 9.2 hr in the same sense as the orbit, in order to produce synchronous rotation. Particle color indicates body of origin.

**Movie 4.**

**Peak accelerations during the merger of spherical components at 2.9 m s$^{-1}$ and an impact angle of 80°.** This is the same simulation as in Movie 3, but now particle colors correspond to the maximum acceleration experienced by each particle up to the time shown. Darkest blues correspond to $3.5 \times 10^{-4}$ m s$^{-2}$ and darkest reds to $8.8 \times 10^{-1}$ m s$^{-2}$ on a linear scale. Although some disturbance is experienced by loose surface particles globally (each sphere is settled individually before they experience each other's gravity), the maximum disturbance during the simulation is concentrated in the narrow contact area, or "neck," between the two bodies.

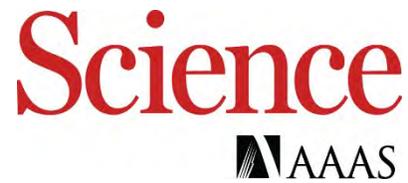

# Supplementary Materials for

## The solar nebula origin of (486958) 2014 Arrokoth, a primordial contact binary in the Kuiper Belt

**Authors:** W. B. McKinnon, D. C. Richardson, J. C. Marohnic, J. T. Keane, W. M. Grundy, D. P. Hamilton, D. Nesvorný, O. M. Umurhan, T. R. Lauer, K. N. Singer, S. A. Stern, H. A. Weaver, J. R. Spencer, M. W. Buie, J. M. Moore, J. J. Kavelaars[10], C. M. Lisse[9], X. Mao[1], A. H. Parker[5], S. B. Porter[5], M. R. Showalter[7], C. B. Olkin[5], D. P. Cruikshank[6], H. A. Elliott[11,12], G. R. Gladstone[11], J. Wm. Parker[5], A. J. Verbiscer[13], L. A. Young[5], and the New Horizons Science Team.

Correspondence to: mckinnon@wustl.edu

**This PDF file includes:**

**New Horizons Science Team (Co-Investigators for the Arrokoth Encounter)**

S. Alan Stern, Southwest Research Institute, Boulder, CO, USA.

Harold A. Weaver, Johns Hopkins University Applied Physics Laboratory, Laurel, MD, USA.

Catherine B. Olkin, Southwest Research Institute, Boulder, CO, USA.

John R. Spencer, Southwest Research Institute, Boulder, CO, USA.

J. Wm. Parker, Southwest Research Institute, Boulder, CO, USA.

Anne Verbiscer, University of Virginia, Charlottesville, VA, USA.

Richard P. Binzel, Massachusetts Institute of Technology, Cambridge, MA, USA.

Daniel T. Britt, University of Central Florida, Orlando, FL, USA.

Marc W. Buie, Southwest Research Institute, Boulder, CO, USA.

Bonnie J. Buratti, Jet Propulsion Laboratory, California Institute of Technology, Pasadena, CA, USA.

Andrew F. Cheng, Johns Hopkins University Applied Physics Laboratory, Laurel, MD, USA.

Dale P. Cruikshank, NASA Ames Research Center, Moffett Field, CA, USA.

Heather A. Elliot, Southwest Research Institute, San Antonio, TX, USA.

G. Randall Gladstone, Southwest Research Institute, San Antonio, TX, USA.

William M. Grundy, Lowell Observatory, Flagstaff, AZ, USA.

Matthew E. Hill, Johns Hopkins University Applied Physics Laboratory, Laurel, MD, USA.

Mihaly Horanyi, University of Colorado, Boulder, CO, USA.

Don E. Jennings, NASA Goddard Space Flight Center, Greenbelt, MD, USA.

J. J. Kavelaars, National Research Council of Canada, Victoria, BC, Canada.

Ivan R. Linscott, Stanford University, Stanford, CA, USA.

Jeffrey M. Moore, NASA Ames Research Center, Moffett Field, CA, USA.

David J. McComas, Princeton Plasma Physics Laboratory, Princeton University, Princeton, NJ, USA.

William B. McKinnon, Washington University in St. Louis, St. Louis, MO, USA.

Ralph L. McNutt, Johns Hopkins University Applied Physics Laboratory, Laurel, MD, USA.

Alex H. Parker, Southwest Research Institute, Boulder, CO, USA.

Simon B. Porter, Southwest Research Institute, Boulder, CO, USA.

Silvia Protopapa, Southwest Research Institute, Boulder, CO, USA.

Dennis C. Reuter, NASA Goddard Space Flight Center, Greenbelt, MD, USA.

Paul M. Schenk, Lunar and Planetary Institute, Houston, TX, USA.

Mark R. Showalter, SETI Institute, Mountain View, CA, USA.

Kelsi N. Singer, Southwest Research Institute, Boulder, CO, USA.

Leslie A. Young, Southwest Research Institute, Boulder, CO, USA.

Amanda M. Zangari, Southwest Research Institute, Boulder, CO, USA.

**Materials and Methods**

To model 2-body impacts and potential mergers of granular aggregates, we use PKDGRAV, an N-body code with an implementation of the soft-sphere discrete element method (SSDEM) for collisions between spherical particles. PKDGRAV uses the *k*-d hierarchical tree algorithm to reduce the computational cost of calculating interparticle forces and runs in parallel to reduce the time necessary to perform simulations with large numbers of particles (*41*). SSDEM allows particles to interpenetrate, with restoring forces implemented as springs with a user-adjustable spring constant. The implementation of SSDEM in PKDGRAV has been described in detail (*42*), with implementation of static, rolling, and twisting friction (*101*) and interparticle cohesion (*102*). We only applied cohesive forces between particles of the same progenitor body, i.e., any contact between a particle from one body and a particle from another is treated as cohesionless. This choice was motivated by initial simulations of the binary merger with cohesion included. After an initial contact between bodies, the size of the neck would continue to grow as particles near the contact point stuck together and pulled others along with them. Because we wanted to use cohesion to capture the effect of material strength, we judged this behavior to be unphysical and so adjusted to the model.

We model each lobe of the present contact binary system separately, under the assumption that they formed separately and merged at some point in the past. We generate spherical "rubble piles" out of many smaller particles. We use approximately 135,000 equal-size particles to model LL (diameter 17.94 km) and 63,000 to model SL (diameter 13.64 km); the sizes of lobes were based on preliminary estimates (*7*). Particle radii are normally distributed with a mean radius of 136 m and a standard deviation of 27 m. Upper and lower radius cutoffs are 163 m and 109 m, respectively. After generating LL and SL we run simulations with each body separately to allow the particles to settle into an equilibrium between self-gravitation and repulsive contact forces. Particles are either frictionless or given gravel-like friction parameters—a static friction coefficient of 1.0, a rolling friction coefficient of 1.05, a twisting friction coefficient of 1.3, and a shape parameter of 0.5; the friction parameters mimic the shear strength of irregular particle shapes in contact (*101*). The normal and tangential coefficients of restitution are 0.2. Simulations of direct collisions (Figs. 4A,B) assumed no rotation of the individual bodies before contact, whereas that for Fig. 4C assumed a synchronous rotation rate appropriate to the density; the latter was done to better simulate the possible final merger conditions of a co-orbiting binary.

**Supplementary Text**

Accretion by hierarchical coagulation

The cold classical Kuiper belt population might have accreted by traditional hierarchical coagulation (*27*). CCKB object formation through a variant of HC has been shown to be viable even in a low-mass (or "light") planetesimal disk (of $\sim 0.1 M_\oplus$), but only if a pre-existing seed population of $\sim 1$-km-scale planetesimals is invoked while simultaneously most of the mass is in cm-sized pebbles or smaller (*28*). Formation of km and sub-km scale planetesimals has not been demonstrated for traditional HC models, which was one of the motivations for the development of SI models (e.g., *57*). Models (*28*) typically take 10s to 100s of Myr to accrete the larger bodies of the CCKB, ignore possible dynamical stirring by Neptune, and yield a CCKB that is too massive today (see below).

Characteristic planetesimal mass from the streaming instability

Numerical simulations have focused on the formation of larger ($\sim 50$-$100$ km scale) planetesimals in the Kuiper belt, i.e., those at the break in the observed KBO size-frequency distribution (e.g., *24*). We focus on the implications of the SI for a lower-mass, outer protoplanetary disk, i.e., the cold classical region. The gravitational mass scale ($M_G$) in the streaming instability is given, in the CCKB object region, by $4\pi^5 G^2 \Sigma_p^3 / \Omega_K^4 \sim 10^{14}$ kg $\times (\Sigma_p / 0.016$ kg m$^{-2})^3$, where $\Sigma_p$ is the surface mass density in pebbles, $\Omega_K$ is the heliocentric orbital frequency, and $G$ is the gravitational constant (*26*). If we adopt $\Sigma_p = 0.016$ kg m$^{-2}$ from (*28*), who spread $0.1 M_\oplus$ of solids between 42 and 48 au to form their "light disk," the characteristic planetesimal mass produced by an SI-induced GI would be similar to the mass of Arrokoth ($\sim 10^{15}$ kg), given the sensitivity of $M_G$ to $\Sigma_p$. In comparison, to produce a characteristic CCKB object diameter of $\sim 100$ km ($M_G \sim 3 \times 10^{17}$ kg) (*103*) requires an order of magnitude more solid mass in pebbles. Neither of these total masses, as planetesimals, violate the surface mass constraint (*20*) for halting Neptune's outward migration.

Because of the limited dynamical excitation of the cold population, (*19*) have argued for modest dynamical depletion of the CCKB at most since its formation, perhaps by no more than a factor of ~2. This implies that the mass of sizeable objects in the cold classical region has always been low. Given the very low mass of CCKB objects today (~$0.001 M_{\oplus}$ (*103*)), this suggests that the formation of CCKB objects via the streaming (or other) instability —from a larger reservoir of solids that would allow larger bodies than Arrokoth to accrete—must have been intermittent in space and time or otherwise inefficient during the lifetime of the protosolar nebula (*104*). The alternative, that the CCKB was originally more massive and lost substantial mass to collisional grinding (*105*), is not consistent with the lack of evidence for collisional processing of Arrokoth (*7, 8*) and the large fraction of loosely bound binaries among the cold classicals (e.g., *106*). Sporadic or inefficient planetesimal formation could be related to a globally lower pebble/gas ratio owing to gas drag drift of pebbles (*92, 94*), but with local pebble concentrations due to zonal flows or other mechanisms (e.g., *107*).

Alternative particle concentration mechanisms to the streaming instability

In addition to SI, nebular turbulence likely led to particle concentrations at corresponding eddy scales (*21*), but whether such concentrations led to GI and planetesimal formation has received less attention (*108, 109*). SI is a dynamic particle concentration mechanism, which is expected to occur over a range of protoplanetary disk conditions and pebble sizes, possibly in tandem with other particle concentration mechanisms (*57, 110*). For example, the surface mass density of gas in the outermost protosolar nebula, and of the CCKB object formation zone in particular, plausibly should have been low enough for cosmic-ray and x-ray induced ionization and active magneto-rotational instability (MRI) (*57*). Levels of turbulence associated with MRI (*111*) may act to suppress SI, at least in its classic laminar form (*112*). MRI might not reach the disk midplane, however, where SI would take place (*57*, *113*), and SI and MRI can act in tandem, with SI enhancing particle concentrations on a smaller scale (*109*, *114*).

## LL and SL as possible Roche ellipsoids

The spin and angular momentum of Arrokoth can be normalized by the critical rotation rate $\omega_c = \sqrt{\frac{4}{3}\pi\rho G}$ and $mR_e^2\omega_c$, respectively, where $m$ and $R_e$ are the total mass and equivalent spherical radius. The normalized spin and angular momentum for Arrokoth are ~0.29 and ~0.36, respectively, assuming a mass ratio of 2:1 for LL and SL and $\rho = 500$ kg m$^{-3}$ for both lobes. These values resemble those for critically stable Roche ellipsoids of the same mass ratio, about 0.28 and 0.26, respectively (see *99*), but the correspondence breaks down for lower densities (the normalized values scale as $\rho^{-1/2}$).

## Shape of Arrokoth's individual lobes

The individual mapped units on LL may indicate the merger or assembly of discrete multi-km-scale planetesimals (*7*, *8*). If so, to create such a lenticular or ellipsoidal body as LL requires that the mergers were themselves not very energetic or high velocity. These velocity conditions would have been met in a collapsing particle cloud (*47*, *58*), but not during heliocentric hierarchical coagulation generally; the latter implies speeds in excess (or greatly in excess) of the escape speed from LL. Low cohesion and a near absence of internal friction would have been necessary mechanically at the time of the LL assembly collisions as well. Otherwise, the shape of LL would much more reflect the shapes of the individual subunits from which it was built (as in Fig. 4B).

Alternately, the LL and SL lobes could have accreted directly in a collapsing rotating particle cloud from myriad small pebbles (*47*), and acquired their lenticular shapes naturally. Arrokoth is a contact binary, and not a single, broadly ellipsoidal body, so the dynamical regime that fostered quasi-equilibrium shapes of the individual lobes must not have been applicable when the two bodies themselves finally merged. This suggests that the merger of two lobes (LL and SL) may not have occurred in the pebble cloud itself, but at some later time after the pebble cloud cleared (on an ~$10^4$ yr time scale (*58*)), when the lobes may have acquired some modest measure of strength (*87*).

Arrokoth's neck appears somewhat bent or tilted in the direction of rotation [(*8*), clockwise in their figure 1A], as if this was due to a final, tangential mass displacement at the contact surface during a merger. Alternately, the bending could be due to some later mechanical failure/distortion at the neck. The edges of LL and SL observed on approach often display linear segments [(*8*), their figure 3], as if the portions of the lobes just out of sight had been sheared off, though this is not clear in the available images. Perhaps these are the outcomes of earlier on-edge, glancing collisions between the lobes (as in Fig. 4A). Alternately, these apparent facets may have been caused by higher-velocity impacts and mass loss, such as have affected the asteroids (*115*), although a heliocentric impact explanation is not consistent with the dearth of large craters on the visible faces of LL and SL, save perhaps for the largest, "Maryland" (*8*).

**Table S1.**

**Estimated YORP coefficients for near-Earth asteroids.** From photometric measurements of asteroid rotational accelerations, an empirical torque coefficient $Y$ can be estimated from the YORP torque equation $\frac{d\omega}{dt} = \left(\frac{Y}{2\pi\rho R^2}\right)\left(\frac{L_\odot}{4\pi c \bar{a}^2}\right)$, where $\omega$, $\rho$, and $R$ are the spin rate, density, and equivalent radius of the asteroid in question, $L_\odot$ is the solar luminosity, $c$ is the speed of light, and $\bar{a} = a_\odot \sqrt[4]{1 - e_\odot^2}$ is the solar-flux-weighted mean heliocentric distance (*77*). In the table $P$ is the rotation period, $\omega/(d\omega/dt)$ is the spin-rate doubling time, and data sources are indicated. For Itokawa and Bennu the densities are known; for the others 1500 kg m$^{-3}$ is assumed, except for (54509) YORP, (1862) Apollo, and (161989) Cacus, which are likely more monolithic and denser (*116, 117, 118*). Note that all these asteroids are spinning up; at present none are spinning down.

| Object | d$\omega$/d$t$ (×10$^{-8}$ rad/d$^2$) | $P$ (hr) | $\bar{a}$ (AU) | $\omega$/(d$\omega$/d$t$) (yr) | $R$ (km) | $\rho$ (kg m$^{-3}$) | $Y$ |
|---|---|---|---|---|---|---|---|
| 54509 YORP (*116*) | 350 ± 35 | 0.203 | 0.987 | 5.8×10$^5$ | 0.06 | 2500 | 0.006 |
| 25143 Itokawa (*117*) | 3.5 ± 0.4 | 12.132 | 1.297 | 9.7×10$^5$ | 0.16 | 1195 | 0.0003 |
| 1620 Geographos (*75*) | 1.2 ± 0.2 | 5.223 | 1.206 | 6.6×10$^6$ | 0.98 | 1500 | 0.005 |
| 1862 Apollo (*117*) | 5.5 ± 1.2 | 16.3 | 1.338 | 4.6×10$^5$ | 0.75 | 2500 | 0.026 |
| 3103 Eger (*118*) | 1.1 ± 0.5 | 15.3 | 1.358 | 2.5×10$^6$ | 0.75 | 1500 | 0.003 |
| 161989 Cacus (*118*) | 1.9 ± 0.3 | 3.755 | 1.110 | 5.8×10$^6$ | 0.5 | 2500 | 0.003 |
| 101955 Bennu (*82*) | 4.6 ± 1.8 | 4.296 | 1.115 | 2.1×10$^6$ | 0.245 | 1190 | 0.0008 |